# Why Model Credibility Isn't Enough:
## - Rethinking Trust in Simulation Architectures -

**Romain Barbedienne** [1)] **Adeline Lanugue** [1)] **Rim Kaddah** [1)] **Julien Silande** [1,2)] **Anthony Levillain** [3)]
**Cedric Leclerc** [4)] **Maxime Hayet** [5)] **Boussaad Soualmi** [1)] **Cristian Maxim** [1)]

*1) IRT SystemX, 2 Bd Thomas Gobert, 91120 Palaiseau, France (E-mail: romain.barbedienne@irt-systemx.fr)*
*2) Keysight Technology, La janais 3 rue Pierre et marie curie, 35131 Chartres-de-Bretagne, France*
*3) OPmobility Alphatech, ZAC du Bois de Plaisance, 214 Av. de la Mare Gessart, 60280 Venette, France*
*4)Renault Technocentre, 1 avenue du Golf, 78280 Guyancourt, France*
*5)Stellantis Green-Campus, 43 Rue Jean Pierre Timbaud, 78300 Poissy, France*

**ABSTRACT**: Credibility of a simulation model is an important topic. Several approaches try to quantify the credibility of simulation. However, models are mostly assembled within a simulation architecture. Can the credibility of a simulation architecture be assessed based on the credibility of the models that comprise it? This paper aims to address this issue by providing an overview of the current state of the art in the field of assembly credibility. It will compare sensitivity analysis techniques, qualitative analysis by experts, explainability in AI, and networks. Finally, an assessment of the proposed approaches, based on criteria such as rigor, generalization, and resource requirements, will reveal the strengths and weaknesses of each approach.

## 1.INTRODUCTION

Designing a system requires ensuring that it operates correctly while guaranteeing safety appropriate for its intended use. In recent years, computer simulation has become widely used in the industrial sector to validate these functional and safety requirements. Indeed, for complex systems, simulation is less expensive than conducting real-life tests. Thereby, simulation leads to anticipate and test a system as soon as possible to enhance it. Moreover, it allows to realize a lot of tests in a short time and quickly improve the system. However, to be pertinent and prevent errors in decision making, these tests should be done in a model which is enough credible and robust.

In the M&S domain, credibility could be defined as the degree of confidence that a stakeholder places in a product for a specific purpose based on established independent evidence. To ensure the credibility of a model, it is necessary to conduct efforts to ensure that the system is properly designed and that its verification and validation have been carried out correctly. The credibility of a simulation model is essential to ensure the right decision. Improving the credibility of a simulation model is an essential step, but it can be very time-consuming. It is therefore necessary to strike the right balance between improving the model's credibility and the time spent on this task.

In practice, for decisions, it is necessary to combine models from different teams. Engineers build model architectures composed of several submodels. These models are connected to one another via interfaces. Each interface contains one or more variables that are exchanged between the submodels. The submodels constituting this architecture have their own credibility. These submodels can sometimes even be delivered by suppliers [1]. In this context, is it possible to evaluate the credibility of a simulation architecture based on the credibility of the models that comprise it?

This paper is a analyze different solution that aimed at evaluating scientific approaches to addressing this issue. The first part of this paper is focus on the methods that allows to evaluate credibility of a simulation model. We will discuss the direct application of these methods to the evaluation of simulation architectures. The second part will explore the methods that can be used to evaluate the impact of a model in a simulation architecture. The third part is an analysis and a comparison of different methods.

## 2.Evaluation of simulation model credibility

The credibility of simulation model is characterized by the degree of confidence that a stakeholder places in a product for a specific purpose based on established independent evidence. Thus, assessing the credibility of a model is subjective; it cannot be measured using precise metrics. However, several studies have focused on rationalizing credibility.

### 2.1.Method for ensuring the credibility of models

NASA has introduced Standard 7009B [2]. This standard defines uniform technical and engineering requirements for the processes, procedures, practices, and methods recognized as standards in modeling and simulation. One of the key elements in assessing modeling and simulation capabilities is the set of recommendations covering the following topics: data traceability, verification, validation, technical review, and product/process management. Each requirement must be met throughout a model's lifecycle, which adds complexity to the design process since the requirements must be reviewed with every modification. This enhances the model's credibility but is particularly suited for high-criticality simulation models.

### 2.2.Frameworks for credibility assessment

Oberkampf et al. [3] introduced the Predictive Capability Maturity Model (PCMM). This framework evaluates the maturity of a simulation model across several dimensions: representation and geometric fidelity, physics and material model fidelity, code verification, solution verification, model validation, and uncertainty quantification and sensitivity analysis. Each dimension is defined for different maturity levels. PCMM allows for a quick assessment of a model's strengths and weaknesses. One limitation of this approach is that it is primarily applicable to 3D models.

Prostep ivip publishes a white paper [4] where they explain the emphasizes the need to standardize the assessment of simulation credibility by incorporating criteria related to quality, accuracy, and criticality, to strengthen stakeholders' confidence in the results obtained. This document introduces a framework for assessing the credibility of processes. Within this framework, the Wheel of V&V and the Credibility Spider Chart help highlight the strengths and weaknesses of a simulation model, enabling stakeholders to quickly assess the model's credibility. This highly promising work needs to be extended to other areas of verification and validation; indeed, the credibility of a model also depends on its design and its specifications.

Finally, IRT SystemX publishes a white paper [5] to improve the interaction between OEM and supplier in the context of simulation. They provide a questionnaire designed to assess the credibility of simulation models across five key areas: robustness and sensitivity, model uncertainty and margin, verification, validation, and use. This questionnaire is summarized in a memo that provides an overview of the level of rigor applied to each model. A minimum score is required for each axis, based on the criticality of the simulation, the system design cycle, and whether or not the model needs to be integrated into a simulation architecture.

### 2.3. Regulations

The European commission has established a regulation aimed at establishing guidelines for the certification of vehicles capable of operating without human intervention [6]. The four parts concern the credibility assessment when using virtual chains tools to validate vehicle. Some rules are linked with the experiences of the team, others about proofs, and the traceability in a documentation. To respect these rules, as soon as possible, in the V-Model enhance the credibility of the model and the system.

### 2.4. Comparison and discussion

The five approaches have the common objective of rationalizing credibility assessment but differ in scope and rigor. PCMM and NASA-STD-7009B are the most scientifically grounded, addressing respectively the technical foundations of a model. The Prostep ivip and IRT SystemX frameworks adopt a more pragmatic stance, offering structured tools that facilitate communication between stakeholders in industrial contexts. However, these approaches remain largely declarative: credibility relies on self-reported information, with limited mechanisms to independently verify the claims made — a significant weakness when models are delivered by external suppliers. The European Commission regulation provides a useful normative baseline but stays general and qualitative, oriented toward certification compliance rather than scientific rigor.

A critical limitation shared by all five frameworks is that they are designed to assess standalone simulation models, and their extension to a simulation architecture raises unresolved scientific challenges. Most validation methods require access to experimental reference data, which is often unavailable at the architecture level — particularly in early design phases or in OEM–supplier relationships where test data cannot be shared. Furthermore, none of these frameworks provides a systematic method to address compositional aspects: even if each submodel is individually credible, the propagation of uncertainties across interfaces and emergent behaviors arising from model coupling remain largely uncharacterized. Addressing these issues requires developing methods capable of aggregating submodel credibility and reducing reliance on declarative assessments in favor of more objective, verifiable evidence. In order to assess the credibility of a model assembly based on its submodels, the remainder of this paper will focus on methods for evaluating the impact of a simulation model within a model assembly.

## 3.Evaluation of impact of a model in an Assembly

### 3.1.Problem statement

A simulation architecture is composed of submodels connected through interfaces, each submodel carrying its own credibility level Figure 1. Even when each submodel is found to be sufficiently credible individually, this does not imply that the whole assembly is credible, as allowable errors from submodels may accumulate to unacceptable levels at the system level. Rather than attempting a direct compositional aggregation of model credibility an alternative strategy consists in evaluating the impact of each submodel on the architecture's output variables of interest. This impact-based approach reframes the problem: instead of asking whether a submodel is credible in isolation, it asks how much a submodel's uncertainty or error propagates to the decision-relevant outputs of the architecture.

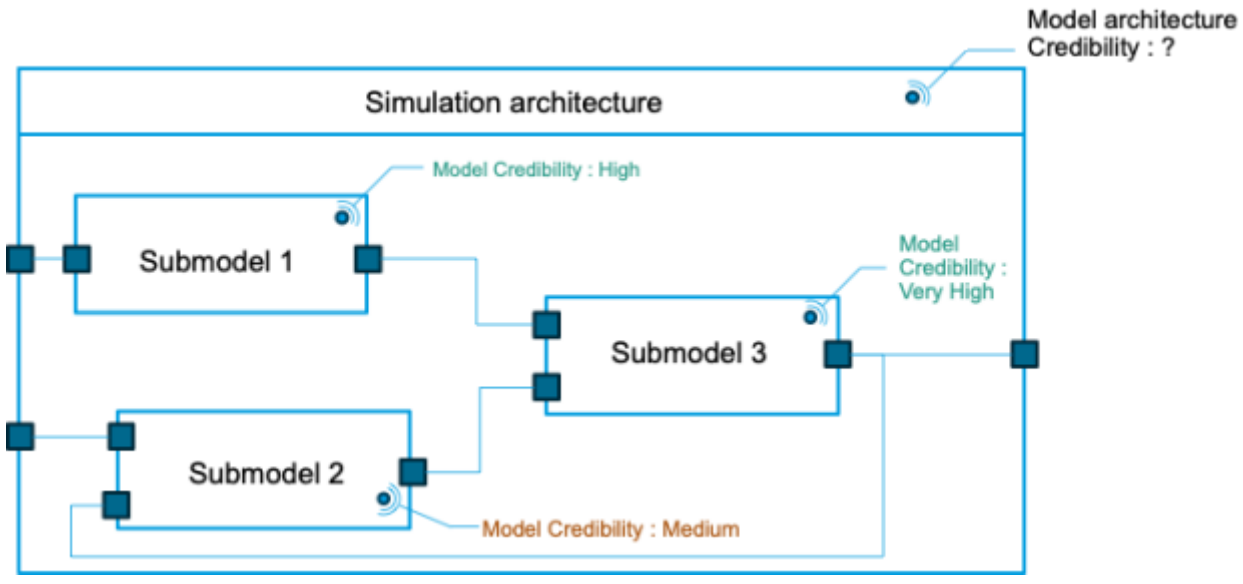

Figure 1 Credibility of simulation architecture

This analysis is inherently context dependent. The validity of a model is defined over a specific domain of model form, inputs, parameters, and responses, which effectively limits its use to the particular application for which it was validated. Thus, the contribution of a given submodel to the architecture's output depends on the scenarios and parameter sets used for decision-making. Two submodels may have very different impacts depending on the operating conditions simulated. This implies that the impact analysis must be conducted specifically for the scenarios and parameter ranges used for the decision making. Furthermore, a model may be valid for one set of experimental conditions and invalid in another; therefore, during the impact assessment, input variables and parameters of each submodel must remain strictly within their respective validation domains to ensure that the conclusions drawn are meaningful. Finally, one last consideration to keep in mind is that simulation models are often compiled in the Functional Mock-up Interface (FMI) format. This facilitates co-simulation, but also allows models to be exchanged between suppliers and OEMs without the risk of sharing proprietary know-how.

To systematically evaluate the contribution of each submodel to the architecture's outputs, two complementary families of methods will be explored in this paper. Sensitivity analysis methods are useful for testing the robustness of results, identifying model inputs that cause significant impact in the output. Moreover, statistical methods offer a complementary perspective by characterizing the probabilistic distribution of outputs and quantifying uncertainty propagation across the architecture. These two approaches will be explored in the following sections.

### 3.2.Sensitivity analysis

Sensitivity analysis is a broad family of methods aimed at determining how variations in model inputs or parameters propagate to model outputs. Depending on whether the analysis is conducted at a single operating point or across the entire parameter space, these methods are generally classified as either local or global. The following subsections review both approaches, discussing their respective theoretical foundations, practical applicability, and limitations in the context of simulation model assemblies.

#### 3.2.1. Local sensitivity analysis

Local sensitivity analysis characterizes how small perturbations in model parameters affect model outputs in the vicinity of a nominal operating point [7]. Formally, given a model $F: \mathbb{R}^p \to \mathbb{R}^q$ mapping a parameter vector $\Lambda \in \mathbb{R}^p$ to an output vector $Y \in \mathbb{R}^q$, the local sensitivity is defined as the partial derivative of model output $y_i$ related to parameter $\lambda_j$ evaluated at a nominal point $\Lambda_1$ (where $i \in [\![1, q]\!]$ and $j \in [\![1, p]\!]$):

$$\mathrm{S_{ij}} = \frac{\partial y_i}{\partial \lambda_j}\Big|_{\Lambda=\Lambda_1}$$

In practice, when analytical derivatives are unavailable, this quantity is approximated numerically using finite difference schemes [8].

From a computational standpoint, local sensitivity analysis is highly efficient. For a model with $p$ parameters, a full sensitivity matrix requires only $p + 1$ model evaluations using a forward difference scheme, or at most 2p evaluations with a centered difference scheme. This makes the approach tractable even for models with a large number of parameters, provided each evaluation is not prohibitively expensive. When derivatives can be computed analytically or via automatic differentiation, the cost reduces further, requiring only a single forward pass through the model alongside its linearization.

However, local sensitivity analysis rests on a fundamental assumption: that the model behavior is approximately linear in the neighborhood of the nominal point $\Lambda_1$. This assumption carries several important consequences. First, the method is strictly valid only when parameter variations remain small relative to the scale over which the model's behavior changes. Second, and more critically, local sensitivity indices are entirely dependent on the choice of nominal point, meaning that the conclusions drawn may not be representative of the model's behavior across its full parameter range. In the presence of nonlinearities, the sensitivity structure can vary dramatically across the parameter space, rendering point-wise estimates potentially misleading. Another example of limitation of local sensitivity is the failure to distinguish between a local minimum and an inflection point. This difference is, however, crucial for determining the robustness of a model Figure 2.

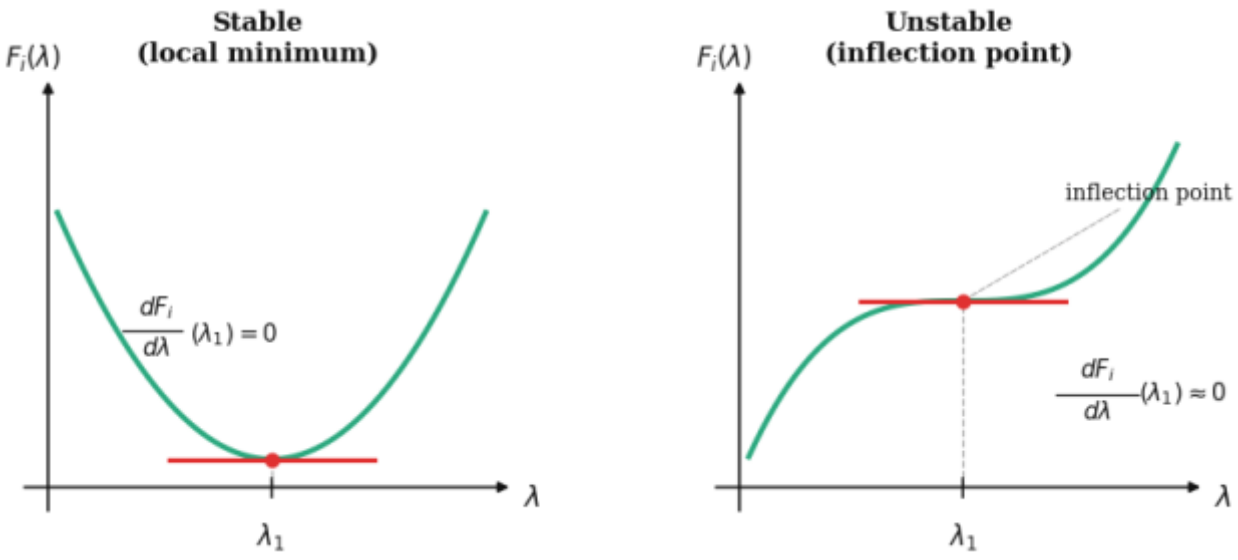

*Figure 2 : Both cases yield* $\frac{dF_i}{d\lambda}(\lambda_1\ ) \approx\ 0$ *so local analysis alone cannot distinguish them*

### 3.2.2. Global sensitivity analysis

Unlike local sensitivity analysis, which characterizes model behavior at a single nominal point, global sensitivity analysis evaluates the influence of parameters over their entire admissible range. By exploring the full parameter space rather than a local neighborhood, this family of methods provides a more robust and representative characterization of model behavior, in particular when nonlinearities or parameter interactions are present. Global sensitivity analysis methods are generally organized into three main families: screening-based approaches, variance-based approaches, and metamodel-based approaches

**Screening-based approaches.** The objective of screening methods is to rank parameters by order of influence at low computational cost, typically as a preliminary step before a more expensive analysis. The most widely used method in this family is the Morris method [9]. The Morris method proceeds as follows. Starting from a randomly sampled initial configuration in the parameter space, each parameter is perturbed individually by a fixed step, and the corresponding elementary effect (defined as the ratio of the output variation to the parameter perturbation) is recorded. This procedure is repeated for r independently sampled configurations, yielding a distribution of r elementary effects per parameter. Two statistics of this distribution are then used as sensitivity measures: the mean of the absolute values estimates the overall influence of the parameter on the output, while the standard deviation serves as an indicator of nonlinear behavior and interactions with other parameters. The method requires only r(p+1) model evaluations in total, which makes it tractable even for models with a large number of parameters. It is also straightforward to implement for any black-box model. Its principal limitation is that it does not quantify parameter interactions precisely, and its conclusions are qualitative rather than quantitative.

The Morris method is frequently used as a first step to identify the most influential parameters before proceeding to a more resource-intensive analysis. At this stage, it is critical not to underestimate the influence of any parameter, as a false negative (incorrectly classifying an influential parameter as negligible) would propagate through all subsequent analyses. To address this risk, Sohier et al. [10] proposed an adaptive variant of the Morris method in which the number of elementary effect samples is increased selectively for parameters with low estimated influence, thereby reducing the probability of misclassification while controlling the total number of model evaluations.

Two further screening approaches are worth mentioning. One-At-a-Time (OAT) analysis [11] varies each parameter individually across its full range while holding all others fixed, and plots the resulting output response to reveal the form of each parameter's influence (including nonlinearities and threshold effects). It is simpler than the Morris method but shares its inability to detect interactions, and the fraction of parameter space explored shrinks super-exponentially with the number of parameters, making it unsuitable for high-dimensional problems. Sequential bifurcation [12] takes a different approach: parameters are grouped and tested in aggregate, with groups progressively subdivided until individual influential parameters are isolated. It is both effective and efficient, requiring relatively few simulations run, but assumes that the model response can be approximated by a first-order polynomial and that the signs of parameter effects are known a priori, which limits its applicability to FMI models.

**Variance-based approaches**. Variance-based methods provide a quantitative decomposition of output variance into contributions attributable to each parameter and to their interactions. The most rigorous framework in this family is that of Sobol indices [13]. This framework decomposes the variance of the output of the model into fractions that can be attributed to parameters or set of parameters.

Sobol indices yield rigorous, interpretable results but require two conditions: parameters must be mutually independent, and the problem must admit a scalar output. When the model produces time-series outputs, it remains possible to apply Sobol indices to scalar quantities of interest derived from the trajectory, such as the maximum, minimum, or time-averaged value. The principal practical limitation is computational cost: estimating Sobol indices requires a structured experimental design, typically Monte Carlo sampling or Latin Hypercube Sampling, and the required number of model evaluations scales as $O(r(p+2))$ for $p$ parameters, making the analysis potentially expensive for high-dimensional problems. Alternative variance-based methods, such as the Fourier Amplitude Sensitivity Test (FAST) [13] and its extended variant eFAST [13], offer reduced computational cost by estimating sensitivity indices through spectral analysis of the model output.

**Metamodel-based approaches.** When the computational cost of direct sampling is prohibitive, metamodel-based methods construct a surrogate that approximates the original simulator at a fraction of the cost. Sensitivity analysis is then performed on the surrogate rather than on the original model. Gaussian process regression, also known as Kriging [14], is among the most widely used surrogates for this purpose: it provides not only a point prediction but also a variance estimate that quantifies prediction uncertainty. Once the surrogate is calibrated on a limited number of model evaluations, sensitivity indices can be estimated analytically or at negligible additional cost.

**Limitations in the context of model assembly.** Applying global sensitivity analysis in a model assembly context raises two fundamental challenges that go beyond computational cost. First, most sensitivity analysis methods are formulated for scalar or static outputs and are not directly applicable to time-series models; when dealing with dynamic systems, the reduction to scalar metrics necessarily entails a loss of information about the temporal structure of the response.
Second, and more fundamentally, all methods described above quantify the influence of individual parameters on an output variable. This raises a conceptual question: if every parameter of a given sub-model has some influence on the decision variable, does that imply that the sub-model itself is influential? Conversely, what if only a small subset of its parameters drives the output, does the remaining sub-model contribute meaningfully? Standard sensitivity analysis provides no direct answer to these questions, because it operates at the parameter level rather than at the model level. Addressing the contribution of a sub-model as a whole requires a different conceptual framework, one in which the object of analysis is the sub-model rather than its individual parameters. Furthermore, standard sensitivity analysis requires varying parameters across their admissible range, whereas the approach sought here aims to assess sub-model influence while keeping all parameters fixed at their nominal values. These two limitations jointly motivate the exploration of alternative techniques for model assemblies.

### 3.3.Statistical approaches

The study of the influence of models within a model assembly is few addressed in the literature. However, Statistical analysis between different variables can be an applied to evaluate the influence between different models in an assembly.

**Statistical correlation approaches.** Granger introduced a method in the context of economy [15]. The idea is to evaluate if two time series $(X_t, Y_t)$ variable are causal from a statistical point of view (where $t \in [T_0, T_1]$). This method combined a regression analysis, and a statistical test; the F-test. The regression analysis consists of Find $A_{i,k}$ (where $i \in \{1,2\}, k \in [\![1, d]\!]$) such as:

$$Y_{t+1} = \sum_{k=1}^{d} A_{1,k} Y_{t-k} + A_{2,k} X_{t-k}$$

Where $d$ is the depth. Then, If $A_{2,k}$ is predominant in front of $A_{1,k}$ it mean that $Y_{t+1}$ depend mainly on $X_{t-k}$. A F-test is performed to ensure that the variables are related. Granger causality can be applied in the context of model assembly. Each model on the simulation assembly, Granger-type statistical correlation can be assessed between each input and output of the model (**Error! Reference source not found.**).

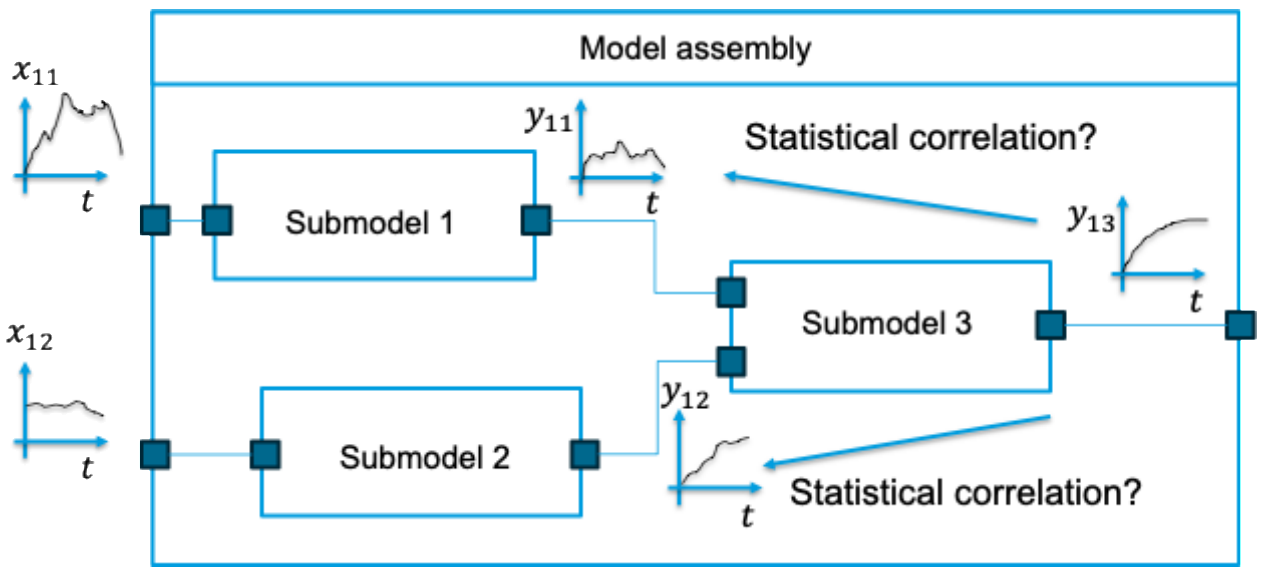

Figure 3 Example of statistical correlation study of submodel3 in a model assembly

The main advantage of this method is than it requires only one simulation to get data, then the granger correlation is applied. Thus, simulation time is really limited. However, the granger causality is adapted for variables linearly correlated. which is not always the case for simulation models.

In the context of regression modeling, Tibshirani introduced lasso (Least Absolute Shrinkage and Selection Operator) regression [16]. This method is a linear regression technique for time series. Let $Y_t$ the variable of interest and $X_{j,t}$ the variables dependent on $Y$ where $t \in [\![T_1, T_2]\!]$, $j \in [\![1, n]\!]$, $n$ is the number of variables dependent on $Y$. Lasso correlation is to find the set $\{\beta_j\}_{j \in [\![1,n]\!]}$ that minimize the function:

$$\min_{\beta_0, \beta_1, \dots, \beta_n} \frac{1}{2} \sum_{t=T_1}^{T_2} \left(y_t - \beta_0 - \sum_{j=1}^{n} \beta_j X_{j,t}\right)^2 + \lambda \sum_{j=0}^{p} |\beta_j|$$

Then the importance of each variable $X_j$ in $Y$ are order by importance of $\beta_j$. This method is particularly interesting because it requires only one simulation. However, Lasso regression is adapted to identify linear dependency between variables.

**Explainability in neural network.** Recent research in the field of explainability in artificial intelligence offers promising approaches to addressing this issue.

Meyes et al introduced the concept of ablation studies in Artificial Neural Networks [17]. The idea of the methodology is to identify the part of a neural network that are the best involve in a decision. The concept is to randomly remove neurons in neural networks to keep the part of neural network which has the greatest influence on the decision variable. This method seems adapted to answer our problematics, because one goal is to identify the subset of models which are the most influential in simulation architecture. But it is not directly applicable, because simulation models cannot directly be removed from a simulation architecture.

**Limitations in the context of model assembly.** Statistical model methods are suitable for simple model, however, linear correlation between different variables of a models can be undetectable by granger causality or lasso correlation. Ablation study seems to be a suitable approach. However, this method can not be adapted to physical model, because in a models assembly, the outputs of a models are linked with input of a downstream model. If a model would be removed, then the inputs of the downstream models will not be connected to the outputs of the upstream models.

## 4.METHODS COMPARISON

### 4.1.Comparison

The preceding sections have reviewed two broad classes of approaches for assessing the credibility of simulation architectures: credibility assessment frameworks applied to individual models (Section 2), and impact evaluation methods — including sensitivity analysis and statistical techniques — that characterize the influence of submodels within an assembly (Section 3). Each approach addresses a different facet of the problem, and none, taken in isolation, provides a complete answer. This section proposes a structured comparison of the seven identified families of methods across four evaluation criteria: objectivity and rigor, cost and execution time, scope and applicability, maturity and standardization.

Objectivity and Rigor evaluate whether the method produces quantitative, reproducible, and verifiable results, as opposed to qualitative or declarative assessments. Cost and Execution Time capture the computational and organizational resources required, including the number of model evaluations and the complexity of the workflow. Scope and Applicability consider the range of models and architectures to which the method can be applied, and in particular whether it extends from individual models to model assemblies. Maturity and Standardization reflect the level of theoretical development, community adoption, and availability of standards or tooling.

Table 1 synthesizes the assessment of each family of approaches against these five criteria. The evaluation is based on the analysis conducted in the previous sections and reflects both the theoretical properties of the methods and their practical constraints when applied to simulation model assemblies.

### 4.2.Analysis

Several observations emerge from this comparison. First, there is a clear trade-off between objectivity and cost. The most rigorous quantitative methods — variance-based global sensitivity analysis (Sobol indices) — provide the highest degree of objectivity, with a mathematically grounded decomposition of output variance, but at a computational cost that scales unfavorably with the number of parameters. Conversely, statistical correlation methods (Granger causality, Lasso regression) require only a single simulation run, making them the most efficient in terms of computational budget, but their restriction to linear dependencies limits the reliability of their conclusions for nonlinear physical models.

The scope of applicability reveals a fundamental gap. All credibility assessment frameworks are designed for standalone models and provide no compositional mechanism for assemblies. Among the impact evaluation methods, sensitivity analysis operates at the parameter level, not at the submodel level, which creates a conceptual mismatch with the problem of assessing submodel influence. Ablation studies from the AI domain address the right conceptual question — identifying the most influential components — but cannot be directly transferred to physical model assemblies due to interface connectivity constraints.

Maturity does not correlate with applicability to the assembly problem. The most mature methods (credibility frameworks, local sensitivity analysis, Sobol indices) are well-established for individual models but were not designed for the compositional question. The most conceptually relevant approach (ablation studies) is the least mature in the simulation context, existing only at a conceptual stage without established methodology or tooling.

Ease of usage which is not a criterion of this table varies significantly but is generally inversely related to methodological depth. Local sensitivity analysis and statistical correlation methods are straightforward to implement for any black-box model, including FMI-compliant components, while metamodel-based approaches and variance-based methods require specialized expertise in experimental design, surrogate modeling, or statistical analysis that may not be available in all industrial teams.

Table 1 Comparison of Approaches with 4 criteria

| Family of Approaches | Objectivity & Rigor | Cost & Execution Time | Scope & Applicability | Maturity & Standardization |
|---|---|---|---|---|
| Model credibility Frameworks (NASA-STD-7009B, PCMM, Prostep ivip, IRT SystemX) | Primarily declarative: credibility relies on self-reported evidence. | Requires systematic documentation and expert review; cost scales with model criticality. | PCMM is mostly limited to 3D physics-based models. Prostep and IRT SystemX are more domain-agnostic. | Well-established due to standardization, industrial use or regulations |
| Local Sensitivity Analysis (Finite Differences, Adjoint Methods) | Provides quantitative, reproducible derivatives. However, conclusions are strictly valid only locally and cannot be generalized across the parameter space. | Requires p+1 to 2p model evaluations for p parameters. Highly efficient, if analytical or automatic differentiation is available. | Valid only in the neighborhood of the nominal operating point. Cannot detect nonlinearities, threshold effects, or parameter interactions. | Well-established mathematical foundations. Widely implemented in all simulation environments. |
| Global SA — Screening (Morris, OAT, Sequential Bifurcation) | Provides a qualitative ranking of parameter influence. Morris statistics (mean and standard deviation of elementary effects) offer reproducible indicators, but results are ordinal rather than quantitative. | Morris requires r(p+1) evaluations. OAT is simpler but explores a vanishing fraction of the parameter space. Sequential bifurcation is efficient but needs prior knowledge of effect signs. | Effective as a preliminary filter for high-dimensional problems. Does not quantify interactions precisely. OAT is unsuitable for high-dimensional systems. Sequential bifurcation assumes near-linearity. | Morris (1991) and OAT are standard tools in sensitivity analysis. Sequential bifurcation is less common but well-documented in the literature. |
| Global SA — Variance-based (Sobol Indices, FAST, eFAST) | Provides a rigorous, quantitative decomposition of output variance into first order and total-order contributions. | Sobol indices demand large sample sizes. FAST/eFAST reduce cost via spectral methods but remain expensive for a lot of parameters. | Applicable to any black-box model with independent parameters. Not directly suited for time-series outputs; requires reduction to scalar quantities of interest. | Sobol, FAST, eFAST is widely used and has been implemented in many tools |
| Global SA — Metamodel-based (Kriging / Gaussian Process Regression) | Objectivity depends on surrogate fidelity. Gaussian processes provide prediction uncertainty, enabling quantification of approximation error. Results are indirect | Initial cost for training data generation, but once calibrated, sensitivity indices can be computed analytically or at negligible marginal cost. Suitable when the original model is computationally expensive. | Applicable to any black-box model. Particularly advantageous for expensive simulators. Accuracy degrades in high-dimensional parameter spaces (curse of dimensionality) unless active learning strategies are used. | Kriging is well-established in geostatistics and computer experiments. Its use for sensitivity analysis is supported by a mature body of literature and robust software implementations. |
| Statistical Correlation (Granger Causality, Lasso Regression) | Provides quantitative, reproducible statistical indicators (F-test for Granger, regularization path for Lasso). However, both methods assume linear relationships, limiting their objectivity for nonlinear models. | Requires only a single simulation run to generate the time-series data. Post-processing is computationally inexpensive. The most efficient family of methods in terms of simulation budget. | Effective for detecting linear dependencies between interface variables. Fails to capture nonlinear couplings, which are common in physical simulation assemblies. | Granger causality (1969) and Lasso (1996) are well-established in econometrics and statistics. Their application to simulation model assemblies is recent and not yet standardized. |
| Explainability in AI (Ablation Studies in Neural Networks) | The ablation protocol is reproducible, but the analogy between neurons and simulation submodels is imperfect. Removing a neuron is well-defined; removing a physical submodel from an assembly disrupts interface connectivity. | Depends on the number of ablation experiments required. In neural networks, cost is manageable; for simulation assemblies, each ablation scenario requires architectural reconfiguration and re-execution. | The idea of identifying the most influential components is directly relevant to model assemblies. However, direct application is not feasible: removing a submodel breaks interface connections, making the assembly undefined. | Ablation studies are standard practice in deep learning (since ~2019). Transfer to simulation model assemblies is at a conceptual stage; no established methodology or tooling exists for this context. |

## 5.DISCUSSION AND CONCLUSION

This paper addressed the problem of assessing the credibility of a simulation architecture based on its constituent submodels. The comparison reveals that no single existing method adequately addresses this issue. Credibility frameworks provide a necessary but insufficient foundation: they ensure that individual models meet quality standards but cannot characterize how errors and uncertainties propagate through interfaces and accumulate at the system level. Sensitivity analysis methods offer tools for understanding parameter-level influence, but the step from parameter influence on submodel influence remains an open scientific question.

Among the approaches reviewed, the conceptual framework of ablation studies appears the most aligned with the problem structure: it directly targets the identification of the most influential components within a composed system. However, its transfer from neural networks to simulation architectures is not straightforward, as removing a physical submodel disrupts interface connectivity and renders the assembly undefined. A viable adaptation could consist in substitution strategies, where a submodel is replaced by a simplified or degraded version rather than removed entirely, thereby preserving interface integrity while enabling assessment of the submodel's contribution to the architecture's output.

Ultimately, the credibility of a simulation architecture cannot be reduced to the aggregation of individual model credibility. It requires a dedicated methodological framework that integrates compositional uncertainty propagation, interface characterization, and decision-relevant impact assessment — a framework that, as this review has shown, remains to be fully developed.

## 6.ACKNOWLEDGMENT

This work has been supported by the French government under the "France 2030" program, as part of the SystemX Technological Research Institute within the AFS project.

## 7.REFERENCES